\documentclass{aa}
\usepackage{psfig}

\begin{document}

\author{N.  Shatsky\inst{1,2} \and A.  Tokovinin\inst{3,1}} 
\institute{
Sternberg Astronomical Institute, RUSSIA, Moscow 119899, Universitetskii pr. 13;\\
\email{kolja@sai.msu.ru}
\and
Royal Observatory of Belgium, BELGIUM, Bruxelles B-1180, av. Circulaire 3.
\and
Cerro-Tololo Inter-American Observatory, CHILE, La Serena, Casilla 603  \\
\email{atokovinin@ctio.noao.edu}
}

\offprints{N. Shatsky}
\authorrunning{Shatsky \& Tokovinin} 
\date{Received September 25, 2001 / Accepted ...}

\title{The mass  ratio distribution of  B-type visual binaries 
in the\\ Sco OB2 association
\thanks{Based on observations collected at the European Southern Observatory,
La Silla, Chile (ESO programme 65.H-0179)}
\fnmsep
\thanks{Tables 1, 3 and full version of Table 2 are only available in electronic
form at the CDS via anonymous ftp to cdsarc.u-strasbg.fr (130.79.128.5) or via
http://cdsweb.u-strasbg.fr/Abstract.html}
}
\titlerunning{Mass ratio distribution of B-type binaries in Sco OB2}

\abstract{
The sample of 115 B-type stars  in Sco OB2 association is examined for
existence  of  visual companions  with  ADONIS near-infrared  adaptive
optics system and coronograph in $J$ and $K_s$ bands.  Practically all
components  in the  separation range  $0\farcs3$--$6\farcs4$ ($45-900$
A.U.)  were  detected, with  magnitudes down to  $K=16$.  The  $K$ and
$J\!-\!K$  photometry  of primaries  and  differential photometry  and
astrometry of  96 secondaries are presented.  Ten  secondaries are new
physical  components,  as inferred  from  photometric and  statistical
criteria,  while remaining  are  faint background  stars. After  small
correction for detection incompleteness  and conversion of fluxes into
masses, an unbiased  distribution of the components mass  ratio $q$ is
derived.   The   power  law  $f(q)\propto  q^{-0.5}$   fits  well  the
observations, whereas  a $q^{-1.8}$ distribution  which corresponds to
random pairing of  stars is rejected.  The companion  star fraction is
$0.20\pm0.04$  per decade  of  separation, comparable  to the  highest
measured binary fraction among  low-mass PMS stars and $\sim$1.6 times
higher  than   the  binary  fraction  of  low-mass   dwarfs  in  solar
neighborhood and open clusters in the same separation range.
}

\maketitle

\section{Introduction}

 Binary  star formation  mechanisms represent  an important  but still
poorly understood  part of  star formation.  This  is why  a concerted
effort is actually deployed to fill this gap from both theoretical and
observational sides.

Observationally,  multiplicity statistics  in  stellar populations  of
different environment, age and mass is one of the most important clues
to binary formation.  Considerable  data have been accumulated for old
low-mass solar-type  nearby stars (Duquennoy \&  Mayor, \cite{DM91} --
hereafter DM91), for  pre-main sequence (PMS) stars and  for the stars
of  intermediate  mass  and  age  in open  clusters  (as  reviewed  by
Duch\^ene \cite{Duchene99}).  In contrast, the multiplicity properties
of high-mass stars remain poorly known.  The most recent comprehensive
study  of  B-type  stars  still  seems  to  be  that  of  Abt  et  al.
(\cite{Abt90}),  based  on  the  spectroscopic  data  and  traditional
catalogues  of visual binaries.   The effort  of Brown  \& Verschueren
(\cite{Brown97}) to measure precise  radial velocities of B-type stars
in Sco OB2 has not yet provided the updated statistics of short-period
systems. Here  we study  the binarity and  mass ratio  distribution of
B-type stars in the separation range of 45-900 A.U.

\begin{figure}
\psfig{figure=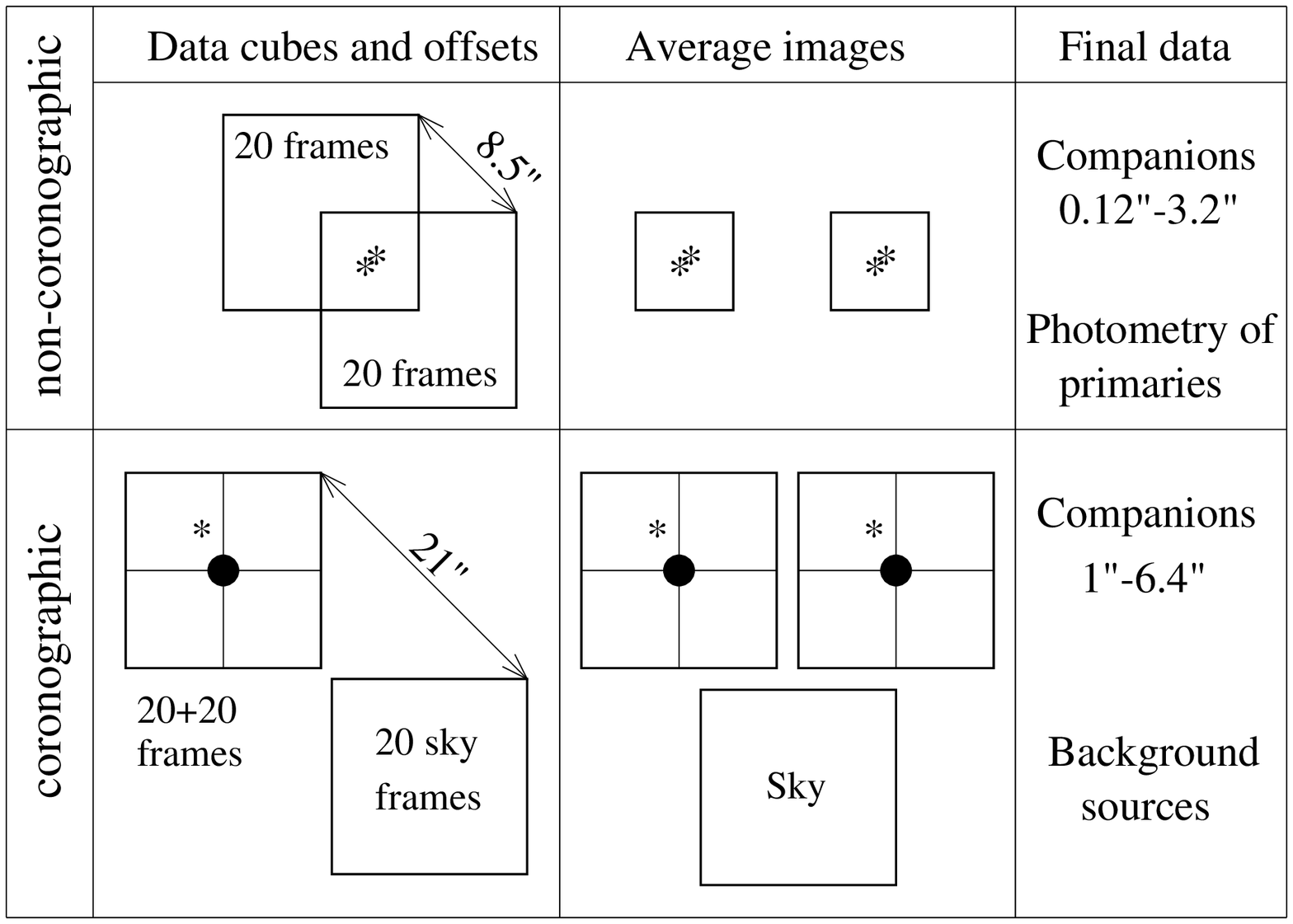,width=8.5cm,angle=-90,clip=t}
\parbox[t]{8.5cm}{
\begin{tabular}{ll}
Telescope & ESO 3.6 m, La Silla, Chile \\
Instrument & ADONIS AO system \\
Camera, detector    & Sharp II + NICMOS3  \\
Image format       & $256^2$ $0\farcs05$ pixels, or  $(12\farcs8)^2$ \\
Coronographic mask   & Radius 1\arcsec  \\
Filters ( $\lambda$/$\Delta\lambda$, $\mu$m): &
J (1.25/0.3), K$_s$ (2.15/0.3) \\ 
One-frame exposure   & $0.05$-$0.4$ s (non-coronographic) \\
    & $3$-$5$ s (coronographic) \\
\end{tabular}
}
\caption{\label{fig:mode}     Scheme    of    ADONIS     (Beuzit    et
al. \protect\cite{Beuzit97}) data and basic instrumental parameters.  }
\end{figure}

Discovery of  massive visual  binaries is limited  to mass  ratios $q$
close  to  1  because  of  the high  intrinsic  brightness  of  B-type
primaries.  Recently, a speckle-interferometric survey of O-type stars
was done  by Mason  et al.  (\cite{Mason98}),  but few new  pairs were
discovered  despite the  increased  angular resolution,  owing to  the
magnitude-difference restrictions  of the optical  interferometry.  On
the other  hand, interferometry in the  infra-red (IR) has  led to the
discovery of 4  additional companions to the 4  brightest stars in the
Orion  Trapezium  (Weigelt  et al.   \cite{Weigelt99}).   S\"oderhjelm
(\cite{Soderhjelm97}) obtained  the unbiased distribution  of the mass
ratio of A and F  binary stars from the magnitude differences measured
by Hipparcos, but  only for $q>0.6$ where detection  was complete. For
A-type  stars, $f(q)$  is roughly  uniform  in this  interval at  short
(60-120  A.U.)  separations and  slightly rises  towards small  $q$ at
larger (240-480 A.U.)  separations.

In  1997  we  used  the  ESO  3.6  m  telescope  with  coronograph  to
investigate the  advantages that Adaptive  Optics (AO) offers  for the
detection and study of  low-mass companions to B-type stars (Tokovinin
et al.   \cite{Tokovinin99} --  TCSB99).  The companion  detection was
complete  in the  separation range  from $1''$  to $6''$  and  for the
magnitude  difference up  to $10^m$  and more  in the  $K$ photometric
band.  High dynamic range imaging and reduced luminosity difference in
the $K$ band (compared to  the visible) give access to companions with
masses  down  to the  bottom  of the  Main  Sequence  and below,  thus
permitting for  the first time  to obtain a complete  companion census
and the  unbiased mass ratio  distribution in the accessible  range of
separations.  The use of AO helps  to reduce the residual wings of the
Point Spread Function (PSF)  outside the coronographic mask, but, more
importantly, concentrates the light from secondary companions into the
diffraction-limited   image  cores,   thus  greatly   improving  their
detectability against  the primary component's wings.   Bouvier et al.
(\cite{Bouvier97}, \cite{Bouvier01}) already used AO to study binarity
in open clusters.

Low-mass binary  companions to B-type stars are  sometimes detected by
their X-ray emission.  This method is complementary to  AO imaging: it
is sensitive  to all  separations but can  not detect  components with
lowest  masses   (see  discussion  in  TCSB99).   Recently  Hubrig  et
al. (\cite{Hubrig2001})  observed a sample of X-ray  selected stars of
late B  spectral type  with AO and  detected new optical  and physical
components.  They did not derive improved binary statistics from these
data.

In this paper we probe for binarity a homogeneous sample of 115 B-type
stars in the Sco OB2  association.  This association is ideally suited
for the studies of B-stars  binary properties for several reasons.  It
is among the  closest to the Sun ($d=145$ pc),  has a well-defined age
with small spread  (from 4 to 15 Myr for  the different sub-groups, de
Zeeuw et  al.  \cite{deZeeuw99}), and is  relatively well investigated
in many  respects (de Geus  et al.  \cite{deGeus89}).  Practically all
B-type  stars  were  observed   by  Hipparcos  which  provided  secure
membership  status   and  additional  constraints   on  binarity  from
astrometry. The binary statistics of  low-mass PMS stars in Sco OB2 is
available for  comparison (K\"ohler et  al.  \cite{Kohler00}, Brandner
et  al.   \cite{Brandner96}).   Thus,  it  is possible  to  check  the
theoretical predictions  about the dependence of  binary statistics on
primary mass.

In Sect.~2 we describe the observing method and the characteristics of
our sample of B-type stars. In Sect.~3 the data processing is outlined
and the  limits of companion detection  are derived. The   $J$ and
$K$ photometry of the known and newly discovered components and of the
primary stars is  given. It is interpreted in  Sect.~4, where the mass
ratio distribution is derived.  The results are discussed and compared
to other works in Sect.~5.

\section{Statistical sample  and observations}

Our  list of  targets is  based on  the work  of Brown  \& Verschueren
(\cite{Brown97}) who provide recent data  for OB-type stars in the Sco
OB2  association. The  membership  of these  stars  in association  is
confirmed by the Hipparcos data (de Zeeuw et al. \cite{deZeeuw99}).  A
few additional targets were also selected from the latter work.

Eight visual  binaries with separations from  $1\arcsec$ to $6\arcsec$
and  small magnitude  difference were  removed from  the observational
program because  they are not  suitable for wave-front  sensing. These
objects were included in  the final statistical analysis, however.  We
presume that there are no  additional companions to these stars in the
studied  separation range, because  most of  such companions  would be
dynamically unstable  (TCSB99).  The basic  data for the  target stars
are  given in  Table~1.  The  interstellar  extinction is
generally small, it is taken from de Geus et al.  (\cite{deGeus89}) or
estimated  from $(B\!-\!V)$ color.   Note that  some pairs  of targets
belong  to the  same wide  multiple systems;  nevertheless,  they were
observed and analyzed independently, as described below.

\smallskip

The observation were performed from 24/25 to 28/29 May, 2000. For each
target star,  we obtained  a sequence of  images (so called  {\em data
cubes})  in $J$  and $K_s$  (hereafter  $K$) filters.   Data with  and
without    coronographic   mask    were   taken    in    each   filter
(Fig.~\ref{fig:mode}):

\begin{enumerate}

\item Short  integration time images ($T_{int}$  limited by saturation
of the  detector) were  taken without coronograph  in  ``1/4
frame'' mode: the image of target star was placed in the center of one
of  the  detector  quadrants for  the  first  half  of the  data  cube
acquisition, and then, with the help of the chopping mirror of ADONIS,
shifted to the opposite quadrant  for the rest of cube.  Binaries with
separations of $0\farcs12$ -- $3\farcs2$  can be observed in this mode,
with partial coverage up to $9\arcsec$.

\item  Long integration  time cubes  were taken  with the  target star
placed in the center of  detector field and hidden by a coronographic
mask of  $1\arcsec$ radius.   In this mode,  the sequence of  {\em sky
frames} was taken  in between of two sequences  of object frames.
The  covered range of  separations is  thus $1\arcsec$  -- $6\farcs4$.
With coronograph, the detection limit at moderate separations is about
$2^m$ -- $3^m$ deeper than in direct images.

\end{enumerate}

The seeing and transparency during this run were variable; the periods
of  photometric conditions covered  only partially  the first  and the
second nights and the whole last night. During the last night, most of
targets with  newly found companions were re-observed  to secure their
photometric   parameters.    For  target   stars   observed  only   in
non-photometric conditions, the $K$  and $J\!-\!K$ values  in
Table~1 are flagged accordingly.

\section{Data processing and results}
\label{sec:datproc}

In this section  we describe the processing of  data cubes, the search
and measurement of companions, and the photometry of both secondary and
primary components.

\subsection{Primary data reduction}
\label{ssec:primred}

For  primary  data  reduction,  we  used  the  {\tt  Eclipse}  package
(Devillard \cite{Devil97})  since it includes the  special utility for
processing  of images  in ADONIS  format.  The  initial steps  of data
processing  were  standard and  included  subtraction  of average  sky
frames (for  the non-coronographic mode  -- subtraction of  the frames
with a  target star in opposite  quadrant), division of  the result by
the flat field, correction for  bad pixels.  Flat fields were taken on
the dusk sky as in TCSB99.  The photometric precision after flat field
division is 1 -- 2\%.

Series of 20  individual images in each data  cube (both coronographic
and direct) were averaged to  produce two independent final images per
cube (Fig.~\ref{fig:mode}),  or a total  of 8 average images  for each
target in two filters and  in two modes.  The subsequent reduction was
done independently  to assure the  reality of the  detected components
and to assess the precision of photometric and position parameters.
While reducing the data cubes, additional  information of two kinds was
also obtained: 
\begin{description}
\item[\bf Sky offset fields.]

To estimate the surface density  of field stars around each target, we
have  processed   separately  the   sky  offset  images   obtained  in
coronographic mode.   The subtraction of  the dark and  background was
done with the  help of {\em cleaned} background  images.  These frames
were  obtained  by  median  averaging  of the  sky  offset  frames  of
different  targets taken with  the same  filter and  integration time,
thus eliminating all  stars.  No flat field division  was performed on
sky images.

\item[\bf Plane-by-plane flux variations.]

As one  of the diagnostics of non-photometric  conditions, we computed
the variation of integral flux  from the target star between different
planes of  the non-coronographic data  cubes. The variances  above 5\%
were treated as clouds signature.
\end{description}

\subsection{Detection  of companions} 
\label{ssec:det}

\begin{figure}
\psfig{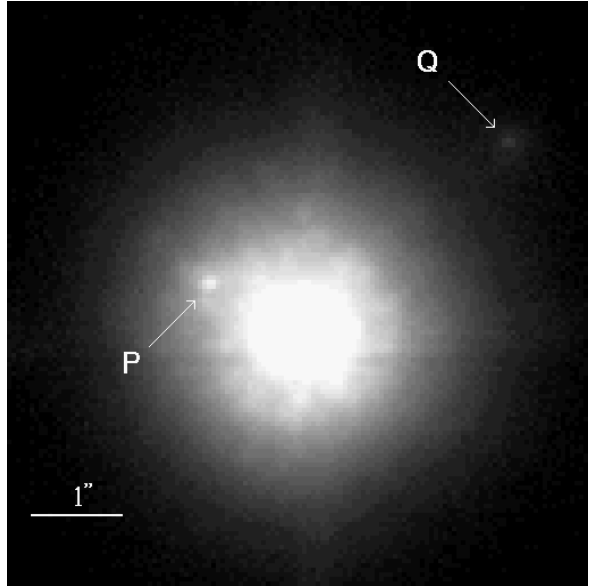}
\caption{\label{fig:hd144987} The non-coronographic image of HD~144987 with two
new companions P and Q. The wings of PSF have the speckle structure.
Primary component is saturated in a given intensity scale.}
\end{figure}

The search  for faint point  sources in the vicinity  of intrinsically
bright B-type  target stars  is a non-trivial  task. The  Point Spread
Function   (PSF)   of   ADONIS    normally   consists   of   a   sharp
(diffraction-limited)  core and extended  wings with  a characteristic
speckle pattern (Fig.~\ref{fig:hd144987}).   This pattern changes from
object  to object  and represents  a major  obstacle for  detection of
faint companions  in AO images (e.g.  Racine  et al. \cite{Racine99}).
We  tried  to reduce  speckle  noise  on  non-coronographic images  by
subtracting PSF models as described in App.~\ref{app:dao} and achieved
noise levels some 1.5--2 times lower.  Speckle structure in the ADONIS
images is  not random (as  assumed in Racine  et al.) but  indeed {\em
semi-static}.

We developed  a special  code {\tt  jupe} (to be  included in  the new
distribution of  {\tt Eclipse}) to determine the  {\em radial profile}
of the  PSF $P(\rho)$ with the  target star hidden  by the coronograph
mask.   First, the  position of  the star  is found  by  comparing the
intensity of PSF wings in $x$ and $y$ directions (cf. TCSB99).  Radial
distance $\rho$ refers to this position.  The profile $P(\rho)$ is the
median value of  intensity at given $\rho$.  We used  the same code to
remove average  radial residuals from  PSF-subtracted non-coronographic
images.

\begin{figure}
\psfig{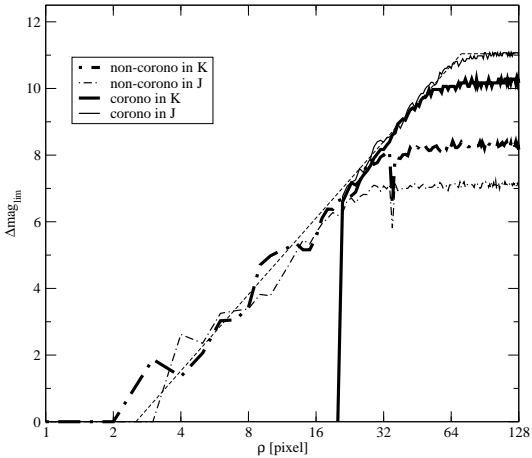}
\caption{\label{fig:detlim} The detection curves for non-coronographic
and    coronographic    images    of    HD~100546    in    $J$    and
$K$ bands.  Coronographic curves  start at  $\rho\!=\!20$  pixels. The
log-linear fit  with a saturation  plateau is used to  approximate the
combined   curve  in   K  (dashed   line).  The   ``spike''  at
$\rho\!=\!35$  pixels is  caused by  subtraction of  the  averaged PSF
fitted by DAOPHOT (App.~\ref{app:dao}).  }
\end{figure}

Images with subtracted $P(\rho)$ were searched for point sources using
the {\tt findobjs} utility  of {\tt Eclipse}.  The detection threshold
was  set  to  $3\sigma(\rho)$,  where  $\sigma(\rho)$ is  the  rms  of
azimuthal intensity fluctuations in the PSF wings at each $\rho$, also
computed  by {\tt  jupe} procedure.   These thresholds  were converted
from pixel  intensities to integrated  fluxes with the help  of Strehl
ratios measured on non-coronographic images.  Thus, limiting magnitude
differences  $\Delta J_{lim}(\rho)$  and  $\Delta K_{lim}(\rho)$  were
derived  for each  frame  (see example  in Fig~\ref{fig:detlim}).  

The  detection   limits  for  the   whole  sample  have   some  common
characteristics.  At small $\rho$, they are approximately proportional
to $R = \log \rho$, with the average slope ${\rm d} K_{lim}/ {\rm d} R
=  1.35 \pm  0.22$.  Further  on, they  saturate at  some  level which
depends  on  the  integration   time.   In  coronographic  images  
$K_{lim}$  saturates  at  about  $K=16.8  \pm  0.5$,  about
$2^m-3^m$ fainter than in non-coronographic images.
It is possible to describe the detection limit by a merged curve which
consists     of    the     non-coronographic    linear     part    for
$\rho\!\le\!1\arcsec$  and  coronographic  part  for  $\rho>1\arcsec$.
This   curve,  typically,   is  continuous   at  the   junction  point
$\rho\!=\!1\arcsec$ (Fig.~\ref{fig:detlim}).   In other words,  in the
area  close  to  the  mask  edge  the  residual  speckle  noise  in  a
coronographic image  with subtracted  $P(\rho)$ and in  the respective
non-coronographic   image   with    subtracted   {\em   average}   PSF
(App.~\ref{app:dao})  is  roughly  the  same.   Individual  log-linear
slopes and saturation levels of the merged detection curves were found
for each target star.   They were subsequently converted into limiting
mass ratios $q_{lim}(\rho)$ (Sec.~\ref{ssec:masses}).

\subsection{Photometry of target stars}
\label{ssec:primptm}

The  majority of  targets  in our  sample  did not  have any  reliable
measurements of  near-IR magnitudes and  colors before our  study.  To
measure the  flux, we must integrate the signal in an aperture
as large as  possible,  reducing the dependence of  the result on image
quality.   Efficiency  of AO  correction  changed significantly  as
seeing varied from $0\farcs5$ to $2\farcs0$ during our run.  Also, the
result  must  be insensitive  to  detector  imperfections  and to  the
presence of other sources in the field of view.  We computed the flux
of the target star as

\begin{equation}
Flux = \sum\limits_{\rho<2\arcsec}I(x,y) + 
   \int\limits_{2\arcsec}^{5\arcsec}{ P(\rho) 2\pi \rho {\rm d}\rho},
\end{equation}
where  the  first term  is  the  sum of  pixel  values  $I(x,y)$ in  a
circle of $2\arcsec$ radius and  the second  term  is an  integral of  the
median  radial  profile.   Median  averaging effectively  removes  all
distant  ($\rho>2\arcsec$) sources and  detector defects.   For binary
targets  with $\rho<2\arcsec$  the  flux of  each  component was  then
computed from the total flux  and the magnitude difference as found by
PSF fitting.

Integration  of the median  profile $P(\rho)$  gives a  somewhat lower
flux than  direct integration of intensity,  because average intensity
of   speckles is higher than their  median intensity. Nevertheless,
special tests have shown that this bias is less than 1\% (or $0.01^m$)
in all  cases.

The  magnitudes  were  reduced   to  zenith  with  average  extinction
coefficients of $0.08$ in $J$ and  $0.10$ in $K$. The errors caused by
extinction  uncertainty  are   negligible  because  all  objects  were
observed close to zenith.  The photometric zero points were determined
from  the primary  standard star  HD~161743 and  confirmed by  5 other
stars for  which $J$ and  $K$ magnitudes were taken from  Simbad. All
determinations are mutually consistent to within $\pm 5\%$.

The  resulting   $K$  magnitudes  and  $J\!-\!K$   colors  of  primary
components are given in Table~1.  The errors reflect the deviations of
individual fluxes  from the mean and  also take into  account the flux
difference    between     two    halves    of     the    data    cubes
(Sect.~\ref{ssec:primred}).      Observations    in    non-photometric
conditions  are  marked by  ``c''  in  the  flags column.   For  these
targets, measured $J$ and $K$ magnitudes represent only upper limits;
the lowest of measured magnitudes was adopted.

We have computed  the expected magnitudes $K_{\rm theo}$  based on the
spectral  types and  visual  magnitudes of  all  sample stars.   These
estimates agree well with the  actual data for the majority of targets
measured  in photometric  conditions:  the difference  $K_{\rm obs}  -
K_{\rm theo}$ shows  a rms scatter of $0.^m15$  and its absolute value
is less than $0.^m4$ for all targets but two.

These two  outliers with $K$ excess  of about $1^m$  are \object{HD~100546} and
\object{HD~143275}.  The first star is  reported to harbor a significant amount
of circumstellar  dust (Augereau  et al.  \cite{Augereau01},  Meeus et
al.   \cite{Meeus01})  which  is  a  natural origin  of  an  increased
infrared luminosity.   The second star ($\delta$  Sco) was intensively
studied  and its  photometry  from Simbad  ($K\!=\!2.75$) agrees  much
better with  $K_{\rm theo}$  than our own  measurement ($K\!=\!1.85$).
Our result could  possibly be explained by an  error of ADONIS shutter
timing at short  (0.02 s) exposure.  HD~143275 is  a brightest star in
our sample.  Nevertheless, other  bright stars were also observed with
such integration time  and an estimated random shutter  error does not
exceed 0.003 s.

On  the other  hand,  $\delta$ Sco  is  a multiple  star with  Be-type
primary and  complex light variations. An extended  study is published
by  Otero  et  al.   (\cite{Otero01})  where  the  observations  of  a
``$\gamma$ Cas-like  outburst'' are  reported.  The peak  magnitude of
$1.^m9$  in  the  visible  was  detected just  two  months  after  our
observations.  The authors  explain this event as being  caused by the
periastron passage in  a close multiple star.  This  is a second, more
attractive astrophysical explanation of our discrepant photometry.

Based on a comparison of observed and expected $K$ magnitudes, we have
extended the validity of companion's photometry to 16 stars which were
observed  on non-photometric  nights  but for  which  $|K_{\rm obs}  -
K_{\rm theo}| <  0.3$ and $J\!-\!K$ colors deviate  from the estimated
by  less than  0.06.  These  cases are  marked as  ``+'' in  the flags
column of Table~1.  The external  errors of our photometry must be not
larger than $\pm 0.^m1$ in $K$ and $\pm0.^m15$ in $J\!-\!K$.

\subsection{Photometry and astrometry of  companions}
\label{ssec:second}

The  program   {\tt  findobjs}  provides   approximate  positions  and
brightness  of  detected  sources.   The final  measurement  of  their
relative coordinates and magnitudes  was made with the profile-fitting
utility  {\tt NSTAR}  of DAOPHOT  package  (Stetson \cite{Stetson87}).
Image of  the primary star  (if single) taken without  coronograph was
selected as PSF model for fitting distant ($\rho>1\farcs75$=35 pixels)
components, whereas  synthetic average PSFs  (App.~\ref{app:dao}) were
used for closer pairs.  Positions of primaries on coronographic images
were   inferred   by  indirect   techniques   with  reduced   accuracy
(App.~\ref{app:ast}).  The characteristic error in position is 0\farcs005 
-- 0\farcs010.

Relative component positions in pixel coordinates were transformed into
arcseconds.   For  calibration, we  used  binary  stars HD~120709  and
HD~199005~AB  measured  by Hipparcos.   Pixel  size  was  found to  be
$0\farcs0497 \pm  0\farcs0005$, and  the orientation of  detector rows
was found  to be  east-west to within  $\pm 0\fdg1$.   Measurements of
several additional known binaries confirmed this calibration.

The magnitudes  and relative positions of 96  secondary components are
given   in  Table~\ref{tab:sec}   (its  full   version   is  published
electronically).

\addtocounter{table}{1}	

\begin{table*}
\caption{
\label{tab:sec}
Photometry and  differential astrometry of companion  stars. Only data
for observed non-optical components  are shown here. Last column gives
the status of companions (see text), new companions are  marked; ``?''
in  the {\em Comp}  column denotes  uncertain detections.  Field {\em
flag} contains ``c'' for data  obtained in non-photometric conditions or ``+''
for     likely     photometric      conditions     (Sect.~\ref{ssec:primptm});
``:'' denotes incomplete resolution.
}
\begin{tabular}{rl|rrrrr|rlrl|cl}
HD & Comp & K & $\epsilon_{\Delta K}$ & 
$J$-$K$ & $\epsilon_{\Delta(J\!-\!K)}$ & flag & 
$\rho\arcsec$ & $\epsilon_\rho\arcsec$ & 
$\theta\degr$ & $\epsilon_\theta\degr$ & $N_{obs}$ & Status \\
\hline
 98718 & B  &  5.86 & 0.01 & -0.03 & 0.01 &    &  0.354 & 0.005 & 143.6 & 0.5 & 1 & P  \\
100841 & P? &  6.81 & 0.27 &       &      & +  &  0.734 & 0.038 & 135.2 & 3.0 & 1 & P   new \\
104878 & B  &  7.00 & 0.02 &  0.07 & 0.08 & c  &  0.698 & 0.008 & 157.9 & 0.2 & 2 & P  \\
108250 & P  & 10.50 & 0.03 &  1.01 & 0.05 & +  &  2.362 & 0.024 &  53.2 & 0.1 & 1 & P   new \\
109668 & P  & 10.94 & 0.26 &  0.82 & 0.29 & c  &  4.853 & 0.049 & 198.3 & 0.1 & 3 & P?  new \\
113703 & P  &  9.16 & 0.02 &  0.47 & 0.02 & c  &  1.551 & 0.016 & 268.2 & 0.2 & 2 & P   new \\
116087 & B  &  7.03 & 0.07 & -0.11 & 0.23 &    &  0.164 & 0.010 & 135.2 & 3.6 & 2 & P  \\
120324 & P  & 10.06 & 0.05 &  1.02 & 0.18 & c  &  4.637 & 0.047 & 304.2 & 0.1 & 2 & P   new \\
120709 & B  &  6.32 &      &  0.02 &      &    &  7.878 & 0.079 & 105.8 & 0.1 & 1 & P  \\
130807 & B  &  6.84 & 0.01 & -0.12 & 0.08 & c: &  0.099 & 0.008 &  86.0 & 7.3 & 1 & P  \\
131120 & P  &  9.43 & 0.10 &  0.85 & 0.12 &    &  1.046 & 0.012 & 161.1 & 0.4 & 2 & P   new \\
132200 & C  &  5.46 & 0.04 & -0.04 & 0.04 &    &  0.128 & 0.008 & 156.4 & 1.9 & 1 & P  \\
132200 & B  &  8.45 & 0.03 &  0.61 & 0.03 &    &  3.950 & 0.040 &  83.0 & 0.1 & 2 & P  \\
133937 & P? & 11.05 & 0.02 &       &      & c  &  0.006 & 0.048 & 293.2 & 0.1 & 1 & P   new \\
136504 & B  &  5.55 & 0.06 &  0.42 & 0.12 & +  &  0.279 & 0.008 & 149.2 & 1.0 & 1 & P  \\
140008 & B  &  9.47 & 0.05 &  0.42 & 0.05 & c  &  0.507 & 0.009 & 132.8 & 0.8 & 1 & P  \\
142378 & B  &  7.78 & 0.01 &  0.23 & 0.02 &    &  0.524 & 0.006 & 119.7 & 0.5 & 1 & P  \\
144217 & B  &  6.80 & 0.05 & -0.78 & 0.12 & c  &  0.292 & 0.010 & 170.5 & 3.1 & 1 & P  \\
144218 & E  &  7.43 & 0.05 &       &      & c: &  0.119 & 0.005 &  36.3 & 2.5 & 1 & P  \\
144987 & P  &  9.75 & 0.08 &  0.57 & 0.12 &    &  1.119 & 0.015 & 116.9 & 0.3 & 5 & P   new \\
144987 & Q  & 12.81 & 0.06 &  1.04 & 0.26 &    &  3.056 & 0.032 & 228.0 & 0.2 & 3 & P?  new \\
145502 & B  &  5.14 & 0.01 &  0.06 & 0.01 & +  &  1.334 & 0.014 &   1.8 & 0.1 & 1 & P  \\
145792 & B  &  8.07 & 0.02 &  0.62 & 0.02 &    &  1.693 & 0.018 & 219.8 & 0.2 & 2 & P  \\
147165 & C  &  4.77 & 0.07 &  0.06 & 0.07 & +  &  0.469 & 0.006 & 244.4 & 0.4 & 1 & P  \\
151890 & P  & 10.33 &      &  2.02 &      & c  &  9.154 & 0.092 & 210.2 & 0.1 & 1 & P?  new \\
157056 & B  &  5.02 & 0.24 &       &      &    &  0.300 & 0.025 & 251.6 & 3.3 & 1 & P  \\

\hline
\end{tabular}
\end{table*}

\addtocounter{table}{1}	

\begin{figure}
\psfig{figure=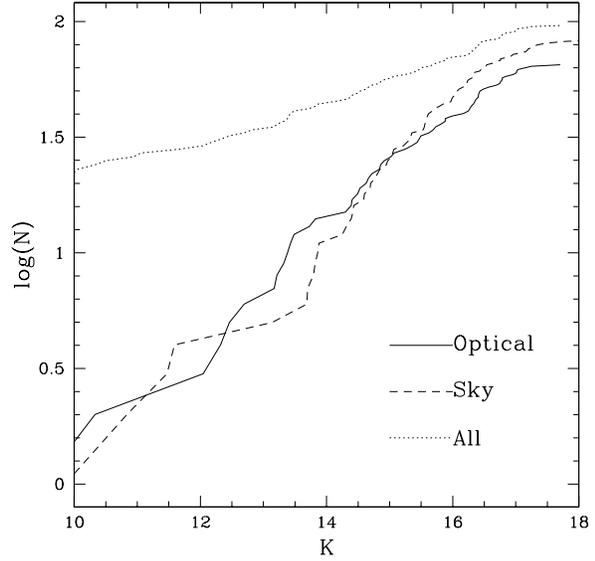,width=8cm,clip=t}
\caption{
\label{fig:srccumdist}
 Cumulative distributions  of $K$ magnitudes  of all sources  found in
target  frames  (dotted line)  and  in  sky  frames (dashed  line).  The
distribution of sources in  target frames which are  considered as
optical  (Sect.~\ref{sec:status})  is  plotted in solid  line  which
resembles the dashed curve of sky-frame background sources.  }
\end{figure}

The approximate  (as given  by {\tt findobjs})  magnitudes of  some one
hundred  field stars  found in  sky  frames are  provided in  Table~3.
These  data are  used to  estimate the  surface density  of background
stellar population.   In both tables the $K$  magnitudes and $J\!-\!K$
colors are given with their  errors.  The errors are inferred from the
scatter  of  individual  values   obtained  from  the  measurement  of
different  frames.    Errors  in  Table~\ref{tab:sec}   are  those  of
magnitude  {\em  differences}  and  do not  include  uncertainties  of
primary  star   magnitudes.   The  cumulative   distributions  of  $K$
magnitudes of sources  around targets and in the  sky frames are shown
in  Fig.~\ref{fig:srccumdist}.   The majority  of  sources are  faint,
close to the detection limits at $K\approx16^m$.

\begin{figure}
\psfig{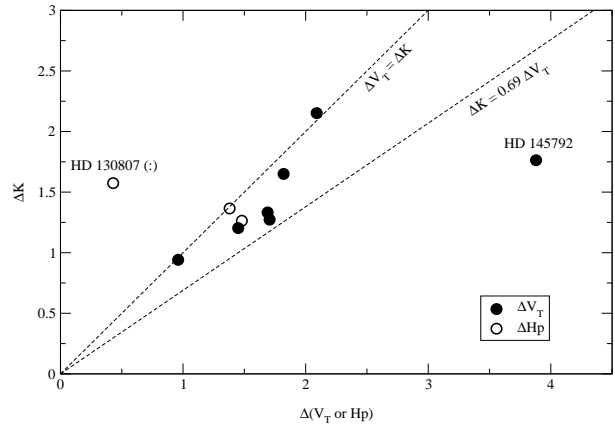}
\caption{
\label{fig:dkvcmp}
Comparison of  the magnitude differences  in the $K$ band  with visual
magnitude difference  from Hipparcos  ($\Delta Hp$) or  Tycho ($\Delta
V_T$, Fabricius \& Makarov \protect\cite{Fabricius00}). The leftmost point 
represents the closest ($\rho\approx0\farcs1$), unreliably measured pair 
HD~130807; the rightmost HD~145792  is possibly an infrared companion. Two
straight lines  represent the equal  differences (upper) and  the ones
expected from the standard relations (Lang \protect\cite{Lang92}).  }
\end{figure}

Few  bright   binaries  from  Table~\ref{tab:sec}   were  measured  by
Hipparcos  or  Tycho  (ESA,  \cite{Hipparcos}) in  the  visible.   The
comparison of their magnitude differences in the $K$ band with $\Delta
V_T$ or $\Delta Hp$ is shown in Fig.~\ref{fig:dkvcmp}.

Four stars  from our sample  (HD 100841, 113902, 126981,  145483) were
also  observed by Hubrig  et al.   (\cite{Hubrig2001}).  They  did not
detect  the  bright companion  to  HD  100841.   The reality  of  this
component  is dubious; it  is not  seen in  our  $J$-band images
either. On the other hand, an additional close (0\farcs2) companion to
HD 145483 was  found by Hubrig et al.; we did  not observe this target
because it has a known companion at $3\farcs8$. Another suspicious component is
HD~133937~P, which is not seen in $J$ but quite prominent in our sole $K$-band
image. This star and HD~100841~P need more observations to confirm their
reality.

\section{Statistics of companions}
\label{sec:interp}

\subsection{The color-magnitude diagram}

\begin{figure}
\psfig{figure=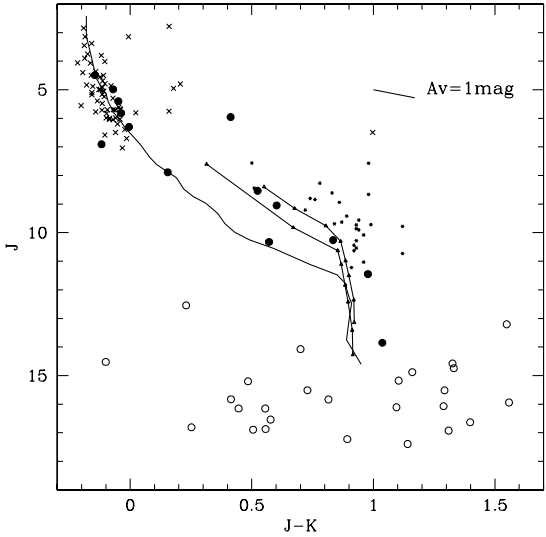,width=8cm}
\caption{
\label{fig:cmd}
The color-magnitude diagram.   Filled circles -- physical secondaries,
empty circles  -- optical secondaries, crosses --  primaries. Only the
stars with valid photometry are plotted. Line is the Main Sequence for
a distance modulus of 5.73, extending  from B0V to M6V.  The 3 Myr and
10 Myr  isochrones for  masses from  1.5 down to  0.08 solar  mass are
plotted as  well (lines with triangles).  Small  asterisks -- low-mass
population   of   the   Sco    OB2   association   from   (Walter   et
al.  \protect\cite{Walter94}).  A  short segment  shows  the reddening
vector for $A_V=1^m$.}
\end{figure}

In Fig.~\ref{fig:cmd}  the color-magnitude diagram is  given. Only the
stars  observed under  photometric conditions  are plotted.   Some red
companions  fall beyond  the  right  boundary of  the  plot.  The  $J$
magnitude is taken as a  photospheric luminosity indicator since it is
known to be relatively free of IR excess, unlike $K$ magnitude. 

The Main Sequence (MS) is traced from the data of Lang (\cite{Lang92})
for a fixed distance modulus  of $5.^m73$ (140 pc). The primaries fall
mostly near MS  or scatter to the right from  MS. The highly deviating
point belongs to HD~100546 (Sect.~\ref{ssec:primptm}). In general, the
measured $J\!-\!K$ colors are validated by this plot.

The isochrones  for 3 Myr  and 10  Myr ages are  based on the  data of
d'Antona \&  Mazzitelli (\cite{dAntona94}, DM94)  (for low-mass stars,
we  used their tracks  computed with  Alexander, Rodgers  and Iglesias
opacities and CM convection, shown  to correspond to real PMS stars by
a number  of authors).  Small triangles  mark the masses  of 1.5, 1.0,
0.7,  0.5,   0.3,  0.2,  0.1,   0.08  solar  masses.    The  effective
temperatures $T_e$  and bolometric luminosities were  converted to the
observed  $(J,J-K)$ parameters.  This  transformation is  not precise,
involving  some  assumptions;  the  tracks themselves  are  not  quite
secure, too.

New tracks were published by a number of authors recently. Among them,
the work of  Baraffe et al. (\cite{Baraffe98}) is  most useful, giving
the synthetic  absolute $J$  and $K$ magnitudes  instead of  $T_e$ and
bolometric  luminosities.  Thus, the  comparison with  observations is
more direct. We found that new  isochrones do not differ from those of
DM94  very significantly.   Although the  slim available  data  on PMS
masses  seem  to  support  the  Baraffe et  al.   tracks  (Steffen  et
al.  \cite{Steffen01}),  the new  tracks  do  not  reproduce well  the
mass-luminosity (M-L) relation of older MS stars.

The actual colors and magnitudes of the low-mass PMS population in Sco
OB2  are plotted in  Fig.~\ref{fig:cmd} by  small asterisks  using the
data of  Walter et al. (\cite{Walter94}). Those  authors estimate that
the masses  of these X-ray selected  stars range from 0.2  to 2 solar,
the typical extinction  is $A_V \approx 0.5$, and the  age is around 1
Myr. We express some reservations  about this age estimate, because OB
stars are much older (de Zeeuw et al. \cite{deZeeuw99}).

\subsection{Status of the secondary components}
\label{sec:status}

The secondaries known  to be physical are either on  MS or above. Some
newly discovered components also fall within the low-mass zone and can
be  classified as  physical.  On  the other  hand, most  of  the faint
secondaries  are optical.   The lower  limit for  physical secondaries
corresponds  roughly  to $J=13$  or  $K=12$  (for  3 Myr  age).   Some
secondaries  have $J-K>1.7$,  falling outside  the graph  boundary. We
presume that they are heavily reddened background stars.

All previously known components  are considered here as physical.  The
reason for this is that the Sco OB2 association has a relatively large
proper motion of 40 mas/yr, and any background component would show up
by its fast motion relative  to primary. This argument does not apply,
however, to the association members that project close to the targets.

All new  components fainter  than $K=12$ or  $J=13$ are  considered as
optical.  The   remaining  bright  opticals  are   identified  on  the
color-magnitude diagram when valid photometry is available. Otherwise, an
uncertain status  is assigned. On the total,  there are 37 physical components,
of which 10 are new (3 of them have uncertain status and 1 is a questionable
star HD~100841~P). The total number of new {\em optical} components is 70.

The status codes in Table~\ref{tab:sec} are: 
\begin{description}
\item[P -] for physical companions;
\item[P? -] for uncertain physical companions;
\item[O -] for  definitely optical (background) companions;
\item[O? -] for likely optical companions.
\end{description}
  The part  of the Table~\ref{tab:sec} reproduced in  this paper gives
information on all observed non-optical companions.  The discrimination
between the optical  and physical secondaries is one  of the important
issues in this study and potentially a weak point.  Naturally, all new
components have small masses and their classification directly affects
the    lower   bin    of    the   mass    ratio   distribution.     In
Sect.~\ref{ssec:qdist}  we  consider  the  3 uncertain  companions  as
physical.  Our guess is  that 1 or 2 of them may  be optical, but this
revision would only reinforce our main conclusions.

\subsection{Statistics of background sources}
\label{ssec:opt}

\begin{figure}
\psfig{figure=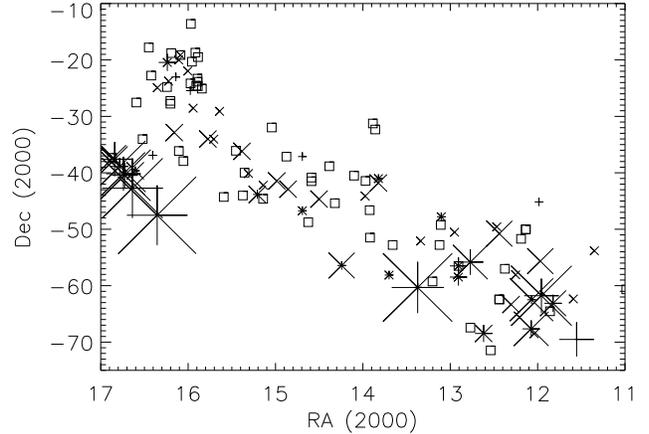,width=8.5cm}
\caption{
\label{fig:opt}
Distribution of the density of background (optical) components in the
sky. The fields with no opticals are plotted as empty squares. The $+$
crosses indicate optical components in the main fields, x crosses --
in the sky fields, with cross size proportional to the 
number of components. 
  }
\end{figure}

In  Fig.~\ref{fig:srccumdist}  the   cumulative  distribution  of  the
optical components  (number of components  $N$ that are  brighter than
given  magnitude)  is  plotted  in  full line.  Only  components  with
$\rho>3\arcsec$ were  selected (at smaller  separations, the companion
detection in the  coronographic frames is affected by  the halo of the
primary).  Dashed line shows the same distribution for the sky fields.
Only  companions  detected  in  {\em  both} $J$  and  $K$  bands  were
selected.

The two curves coincide to within statistical fluctuations, especially
in  the important  region  around $K=12-14$  where the  discrimination
between  optical   and  physical   components  is  critical   for  our
analysis. We  note that  the total surface  of main fields  is smaller
than the  surface of sky fields  by 15\% (central  $3''$ excluded). On
the other  hand, the detection limits  in the sky fields  may be lower
because of the lower Strehl ratio (anisoplanatism).  All in all, there
are 5  components brighter than  $K=12$ in the  sky fields and  3 such
optical components in  the main fields. Taken at  face value, it means
that about 2 components in the main fields may be still mis-classified
as physical.  However, it is  clear that our classification scheme did
not miss a {\em large}  number of faint physical companions, otherwise
we would observe an excess  of ``opticals'' in comparison with the sky
fields.

An important feature of background sources is their highly fluctuating
density  (Fig.~\ref{fig:opt}).  For  45 targets,  no optical
components were  found neither in the  main nor in the  sky fields; on
the  other hand,  in the  remaining fields  a  significant correlation
between the  number of optical components  in the main  and sky fields
was found. In 3 main fields, as much as 7 to 9 optical components were
detected,  with no  less than  6 components  in the  corresponding sky
fields.  This correlation is a  strong argument for the optical nature
of faint components.

Quite often more than one  component identified as optical are present
in the  main fields and have  comparable separations. They  can not be
physical for yet another  reason: non-hierarchical stellar systems are
dynamically unstable and  must disintegrate, given the age  of Sco OB2
group.

\medskip
In  following sub-sections  we consider  the calculation  of  the mass
ratios  $q =  M_2/M_1$. It  will  allow us  to derive  the mass  ratio
distribution $f(q)$ of physical  systems.

\subsection{Estimation of masses and mass ratios}
\label{ssec:masses}

In principle,  masses of  MS stars can  be estimated from  their $J-K$
colors.  However,  the colors are  measured with large errors  and are
distorted by extinction and IR excess. Hence, the best way to estimate
masses of the  companions is to use their  luminosities, preferably in
the $J$  band.  On  the other hand,  the use of  mass-luminosity (M-L)
relation  requires  a knowledge  of  distance  and  age (for  low-mass
stars), it relies also on the yet uncertain PMS tracks.

The  M-L  relations  for  MS   and  PMS  stars  are  taken  from  Lang
(\cite{Lang92}) and Baraffe  et al.  (\cite{Baraffe98}), respectively.
For PMS tracks,  the mass $M$ depends on $A$ (the  logarithm of age in
Myr) and absolute $J_0$ or $K_0$ magnitudes approximately as

\begin{equation}
	\log (M/M_\odot) \approx 0.706 - 0.305 J_0 + 0.545 A ,
\end{equation}

\begin{equation}
	\log (M/M_\odot) \approx 0.440 - 0.313 K_0 + 0.581 A.
\end{equation}
For  MS stars,  the M-L  relation was  approximated by  several linear
segments.   The accuracy of  all these  approximations is  better than
$\pm 10\%$ in  mass.  Of course, the actual  isochrones are not linear
but rather ``saturate''  as the luminosity reaches its  MS value.  So,
the companion  masses were calculated  from both MS and  PMS relations
and  the lowest  of the  two  values was  taken.  The  actual ages  of
subgroups were  used in the calculations: 4.5~Myr  for Upper Scorpius,
14.5~Myr   for   Upper   Centaurus-Lupus   and  11.5~Myr   for   Lower
Centaurus-Crux (de~Zeeuw et al.  \cite{deZeeuw99}).

The M-L  relation is  favorable for mass  estimation: its slope  at MS
roughly corresponds  to $M\propto  L^{0.5}$ and steepens  to $M\propto
L^{0.8}$ for the PMS tracks.  This explains why the mass estimates are
relatively insensitive to data reduction details.

The  masses   of  primary  stars   were  also  estimated   from  their
luminosities rather than from their spectral classes. This was done to
cancel  as much  as possible  the influences  of errors  in distances,
extinction, etc., which affect the masses of primaries and secondaries
in almost the same way and hence have little effect on the mass ratios
$q$.   Even the  errors in  the photometry  caused  by non-photometric
conditions are  compensated to some extent because  $q$ depends mostly
on the magnitude difference.

\begin{figure}
\psfig{figure=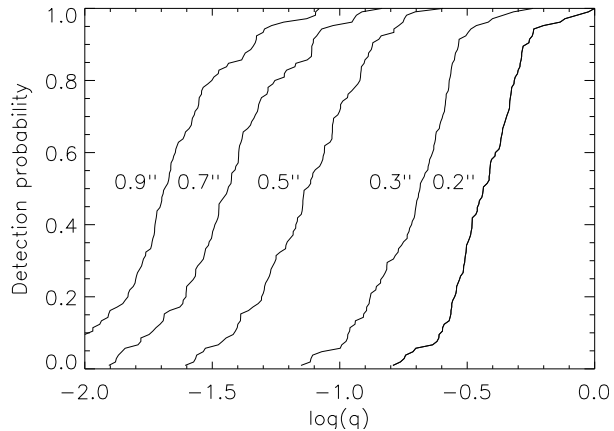,width=8cm,clip=t}
\caption{
\label{fig:bias}
The fraction of detectable components  as a function of mass ratio $q$
for separations from $0\farcs2$ to $0\farcs9$, as indicated by numbers
near the curves.  }
\end{figure}

The detection limits were studied and modeled in Sect.~\ref{ssec:det}.
We convert the derived log-linear relations between $K_{lim}$ and $R =
\log (\rho  /1\arcsec) $  into the limiting  mass ratio  $q_{\rm lim}$
using the M-L-Age relation in the  $K$ band and the actual age of each
target.   $K$-band is used  since low-mass  red companions  are better
detected at longer wavelengths.  These limiting mass ratios are sorted
in increasing  order and  plotted in Fig.~\ref{fig:bias}  as detection
probability  (bias).  If  all frames  were taken  in exactly  the same
conditions,  all  $q_{\rm  lim}$   would  be  identical.   The  actual
distribution  of $q_{\rm lim}$  reflects the  spread in  the observing
conditions, exposure time, target brightness, etc.

The detection  bias is modeled as  a set of linear  functions of $\log
q$,    neglecting   the    ``tails''   of    the    distributions   in
Fig.~\ref{fig:bias}.  Linear models are defined by two parameters, the
$\log  q_{lim}$ where  50\% of  companions are  detected and  the full
range  in  $\log  q_{lim}$.   These  parameters  were  represented  by
quadratic functions of $R$.  The analytical model of detection bias is
plotted in Fig.~\ref{fig:q-rho}.

\subsection{Mass ratio distribution and companion fraction}
\label{ssec:qdist}

\begin{figure}
\psfig{figure=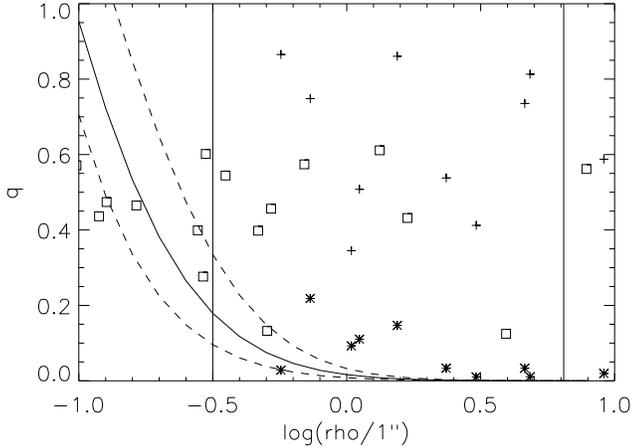,width=8.5cm,clip=t}
\caption{
\label{fig:q-rho}
The  distribution  of the  known  and  measured  (squares), known  and
unmeasured   (pluses)  and   newly  discovered   (asterisks)  physical
components in the $(\log \rho,q)$  plane. The detection model is shown
by  solid  line (50\%  detection)  and  dashed  lines (0\%  and  100\%
detection).  The separation range selected for statistical analysis is
delimited by the vertical lines.  }
\end{figure}

The distribution  of the physical  secondaries in the $(\log  \rho, q)$
plane  is  shown in  Fig.~\ref{fig:q-rho}.  It  is  expected that  the
component  distribution  in $\log  \rho$  should  be uniform  (\"Opik's
law). Indeed, it seems to be  the case. Moreover, there seems to be no
significant correlation between $\rho$ and $q$ in the separation range
studied. This permits to discuss the $q$ distribution for all relevant
separations jointly.

\begin{figure*}
\centerline{
\psfig{figure=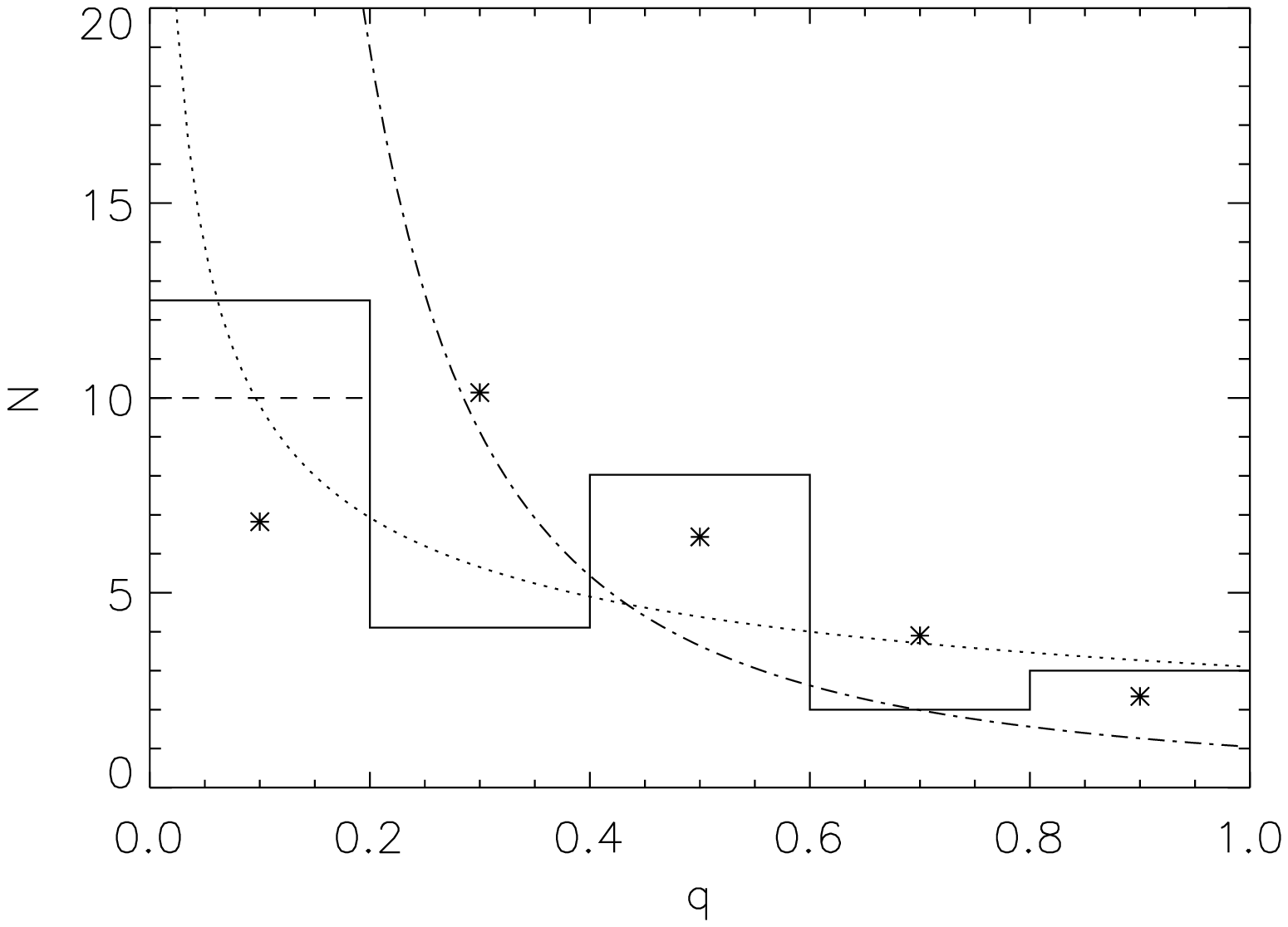,width=8.5cm,clip=t}
\hfill
\psfig{figure=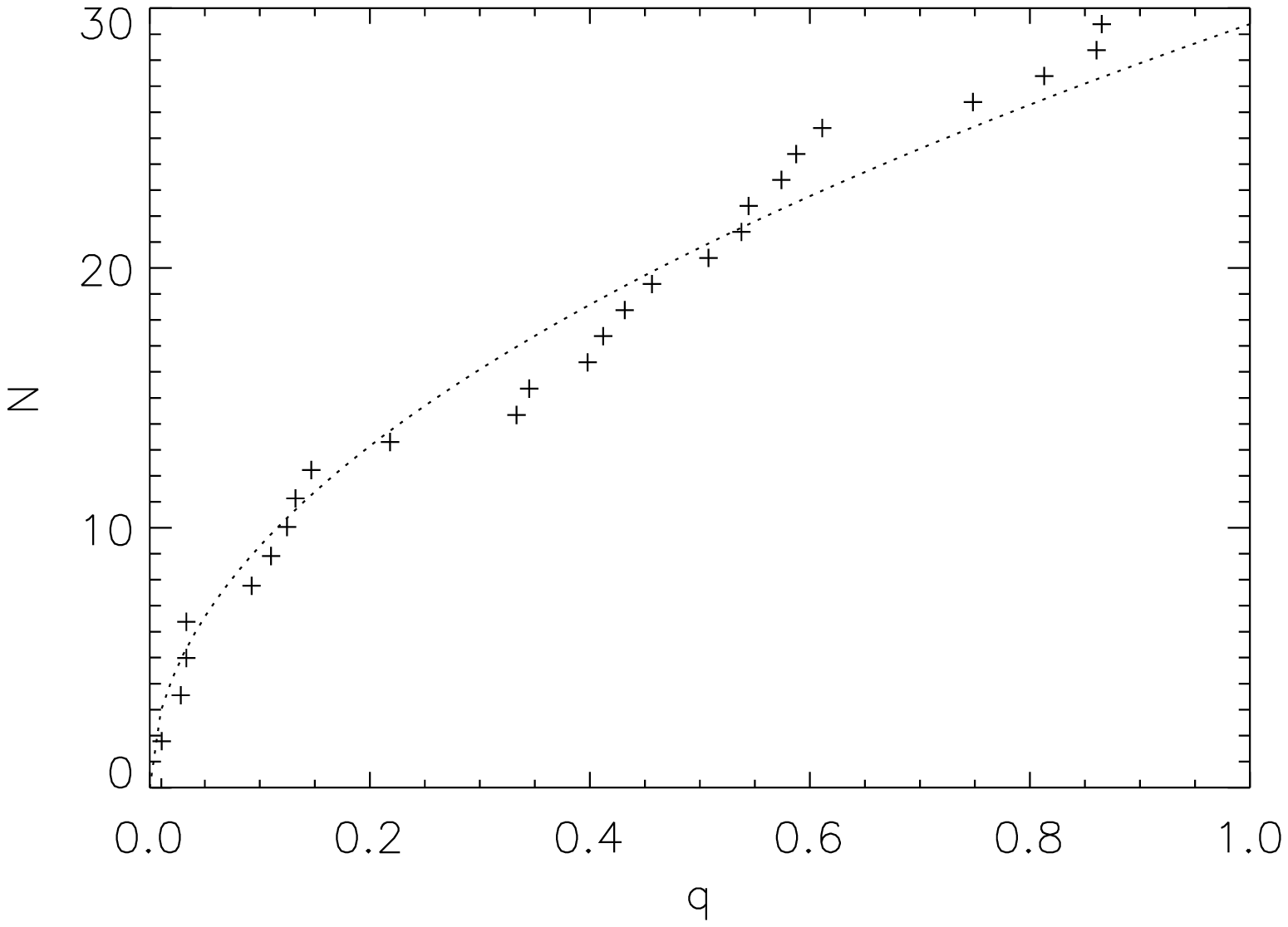,width=8.5cm,clip=t} 
}
\caption{
\label{fig:qhist}
{\bf Left:}  The histogram  of the mass  ratio distribution  $f(q)$ in
separation range from  45 to 900 A.U. (full  line).  Dashed line shows
the  uncorrected value in  the first  bin. Asterisks  -- re-normalized
$f(q)$ of G-type dwarfs with periods $P\!>\!30$~yr ($a>10$ A.U.)  from
DM91.  Dotted line corresponds to $f(q) \propto q^{-0.5}$, dash-dot --
to  $f(q)  \propto q^{-1.8}$.   {\bf  Right:} cumulative  distribution
corrected for  detection incompleteness (crosses) and  its modeling as
$F(q) = q^{0.5}$.  }
\end{figure*}

We  limit  the  statistical  analysis  to the  separation  range  from
$0\farcs3$ to  $6\farcs4$, which corresponds  to 45--900 A.U.   at the
distance of Sco OB2.  For lower separations, the detection bias in $q$
becomes too important.  The upper limit is determined by the half-size
of  the  frames.   Components  at  larger  separations  were  actually
detected in  the corners,  but, as evident  from Fig.~\ref{fig:q-rho},
little can be  gained by extending the limit  to $9\arcsec$ and making
corrections for incomplete surface coverage.

A  total of  27 physical  components fall  in the  selected separation
range  which covers  1.3 decades.  Assuming that  the  distribution in
$\log \rho$ is uniform and that  the distribution in $q$ is smooth, we
estimate  the fraction of  missed components  by integrating  the bias
model  within   the  selected   limits  for  each   bin  of   the  $q$
histogram. The  fraction of detected  components is more than  0.8 for
all  bins,  which means  that  our  incompleteness correction  remains
reasonably small.   The resulting  histogram of $q$  (after correction
for incompleteness) is plotted in Fig.~\ref{fig:qhist} (left).

The same  data reduction steps were  done for the DM94  tracks and two
fixed  ages of  3  and 10  Myr.   The results  are qualitatively  very
similar.  Comparing  the histograms for  these two isochrones,  we saw
that  assuming  a  younger  age  results  in  lower  masses,  slightly
re-distributing the components between the two lowest bins.

In  Fig.~\ref{fig:qhist}  (right) the  same  histogram  is plotted  as
cumulative distribution,  in order to avoid binning.   It is corrected
for detection incompleteness by  increasing the ``weights'' of low-$q$
systems accordingly.  The slope of the cumulative distribution clearly
increases  towards  low  $q$.  Adopting  the  power  law  $f(q)\propto
q^{-\Gamma}$, it  seems that the index $\Gamma\!=\!0.5$  fits well the
data.

The $q$-distribution  does grow towards  low $q$, but only  mildly. On
the other hand,  the power law with index $\Gamma$ from  1.8 to 2.1 is
clearly  rejected. Such  power  law corresponds  to  the initial  mass
function  (IMF)  in  Sco   OB2  (Brown  \cite{Brown98},  Preibisch  \&
Zinnecker  \cite{Preibisch99}, Preibisch  et  al.  \cite{Preibisch01})
and  would apply if  the secondary  components were  selected randomly
from IMF.

The  power-law  distributions   $f(q)  \propto  q^{-\Gamma}$  are  not
integrable  for  $\Gamma >1$,  in  this  case  the total  binarity  is
determined by  the elusive cut-off at  low $q$.   On the contrary,
the actual  distribution is smooth and integrable,  the total binarity
is well defined.  The total number of components (after correction for
incompleteness)  is  29.6 for  the  115  targets  studied and  in  the
separation range of  1.3~dex. This leads to a  companion star fraction
(CSF) of $0.20  \pm 0.04$ per decade of separation  or $0.13 \pm 0.02$
per decade of period.

We repeated the  analysis while excluding the 3  faint components with
uncertain  physical   status  and  the  2   components  with  unsecure
detection.  The  number of  companions in the  $0\farcs 3  - 6\arcsec$
separation   range   becomes    23   (24.85   after   correction   for
incompleteness),  the CSF  is revised  down to  $0.17 \pm  0.04$.  The
lowest bin in the  histogram (Fig..~\ref{fig:qhist} left) becomes 30\%
less, leading to even more uniform $f(q)$ which may be approximated by
a $q^{-0.3}$  law.  This  exercise  shows that  our  conclusions do  not
critically depend  on the remaining uncertainties  in the experimental
data.

\section{Discussion}

Before our  study, we  suspected that the  number of  unknown low-mass
visual components around B-type stars is large, because low-mass stars
are, generally,  much more frequent than high-mass  stars, and because
the  detection  of  such  components  by  traditional  techniques  was
difficult.   Now we  see  that the  newly  detected low-mass  physical
components  are not  so  numerous  and that  the  old detections  were
essentially complete down to at least $q=0.3$. It was indeed necessary
to  go {\em  much}  deeper  in magnitude  difference  to validate  the
historical data!  In this perspective, the fact that most of our newly
detected components are optical is not disappointing.

Our result  is in marked disagreement  with the conclusions  of Abt et
al.  (\cite{Abt90}) who claim  that the distributions of the secondary
components  to B2-B5  stars  follows the  Salpeter  mass function  and
increases  steeply  towards small  $q$  in  the  range of  separations
studied here. Their  analysis is based on the  known visual components
confirmed by  common proper motions.  Still, we  strongly suspect that
most of wide pairs in the B2-B5  sample of Abt et al. are optical.  In
their Table~5  there are 7 trapezium-type systems  with separations in
the $10''$ to $63''$ range and  separation ratio less than 3 which are
likely unstable,  if physical.  When  the spectra of the  components of
116   trapezium-type   systems  were   taken   by   Abt  \&   Corbally
(\cite{Abt00}), they discovered  that only 28 of them  can be physical
-- a proof that most of the cataloged trapezia are indeed spurious.

\begin{figure}
\centerline{\psfig{figure=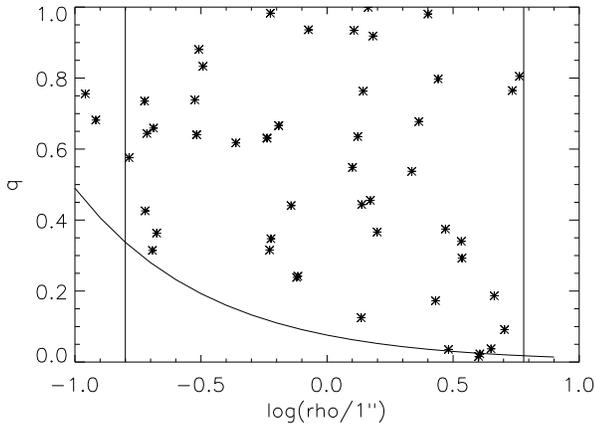,width=8cm}}
\caption{Mass ratio as function of  separation for the 56 low-mass PMS
binaries in Sco OB2 studied  by K\"ohler et al. (\protect\cite{Kohler00}). The
curve  shows indicative  detection bias,  vertical lines  indicate the
separation range of their statistical analysis. }
\label{fig:kohler}
\end{figure}

\begin{table}
\caption{
\label{tab:CSF}
Companion star fraction (number of companions per decade in
separation) in different populations.  
}
\begin{tabular}{|l c ccc|}
\hline
Spectral type,         & $N$ & Range, & CSF &  Ref. \\
environment       &   & A.U.   &      &       \\
\hline
B, Sco OB2       & 115 & 45-900 & 0.20 $\pm$ 0.04  & 1 \\
PMS, Sco OB2     & 118 & 20-900 &  0.21 $\pm$ 0.04 & 2 \\
PMS, Sco-Lup     & 269  & 120-1800 & 0.12 $\pm$ 0.02 & 3 \\
PMS, Tau-Aur     & 104 & 120-1800 & 0.22 $\pm$ 0.04 & 3 \\
G, field         & 164 & 40-900   & 0.12 $\pm$ 0.03 & 4 \\
M, field         & 58  & 10-1000  & 0.11 $\pm$ 0.04 & 5 \\
A-K, Hyades      & 167 & 5-50     & $>$0.16 $\pm$ 0.03 & 6 \\ 
G-K, Pleiades    & 144 & 12-1000  & 0.14 $\pm$ 0.02 & 7 \\ 
G-K, Praesepe    & 149 & 15-600   & 0.15 $\pm$ 0.03 & 8 \\
\hline
\end{tabular}

{\bf References:} 
1 - this work; 
2 - K\"ohler et al. (\cite{Kohler00});
3 -  Brandner et al. (\cite{Brandner96}); 
4 - Duquennoy \& Mayor (\cite{DM91}); 
5 - Fisher \& Marcy (\cite{Fischer92});
6 - Patience et al. (\cite{Patience98}); 
7 - Bouvier et al. (\cite{Bouvier97});
8 - Bouvier et al. (\cite{Bouvier01}). 
\end{table}

In  Table~\ref{tab:CSF}  we  give  a summary  of  statistical  binarity
studies in different  populations, to be compared to  our results. The
sample size $N$ and the  approximate range of separations surveyed are
indicated. The  CSF (fraction of  all companions per unit  interval in
the logarithm  of separation) is  only a weak function  of separation,
hence  it is  legitimate to  compare results  in  different separation
ranges.   For   the  Hyades,  the   CSF  given  by  Patience   et  al.
(\cite{Patience98}) involved  a factor of 2  correction for undetected
systems, hence we preferred  the uncorrected lower limit.

The most  recent study of  the multiplicity of low-mass  population of
Sco  OB2 (K\"ohler  et al.   \cite{Kohler00}) is  very similar  to the
present  work by  the number  of targets  surveyed and  the separation
range ($0\farcs  13 - 6\arcsec$). They  find a CSF of  $0.21 \pm 0.04$
per decade of separation, indistinguishable from our result.  We tried
to process the magnitudes and flux ratios of the 56 systems from their
Tables 2  and 3  in the same  manner as  our data, converting  the $K$
magnitudes into mass ratios with the help of a 3 Myr isochrone and for
the  assumed   distance  of  140   pc.   The  results  are   shown  in
Fig.~\ref{fig:kohler}.   The curve  indicates  our best  guess of  the
detection  threshold,  which is  $\sim  3$  times  higher than  the
threshold given by the authors themselves.  About 7.8 systems in their
sample (mostly  with large separations) were estimated  to be optical.
Taking into  account these  uncertainties, it does  not make  sense to
compare the  histograms of $q$. All that  can be said is  that the $q$
distribution  seems to  be  uniform and  certainly  does not  increase
towards small $q$ as much as would be expected from the IMF slope.

Brandner et al.   (\cite{Brandner96}) provided a comprehensive summary
of the previous  binarity studies among the PMS  stars which were made
in the visible.  The CSF in  the Upper Scorpius and Lupus is $0.12 \pm
0.02$ per  decade of  separation for a  combined sample of  269 stars.
The global CSF  among 525 PMS stars is $0.14  \pm 0.013$, which remains
the  most  statistically  sound  estimate  to date  (however,  in  the
Taurus-Auriga region the CSF is $0.22\pm 0.04$).  Clearly, K\"ohler et
al.    obtained   a   significantly    higher   CSF   for   the   same
population.  However,  Brandner  et  al.   find an  evidence  for  CSF
variations across the Sco OB2,  and the regions studied by K\"ohler et
al. happen to be near the binary-rich zone.  This seems to be the most
plausible explanation of this discrepancy.

Our sample  covers a  large region  in the sky.   For this  reason the
CSF=0.12  measured  by  Brandner  et  al.   is  more  appropriate  for
comparison with our result,  CSF=0.20. Thus, more massive B-type stars
do have an an increased CSF  with respect to the lower-mass PMS stars.
The  same conclusion is reached  by comparing  our result  to the
binary  fraction  of  low-mass   field  dwarfs  and  low-mass  cluster
population (Table~\ref{tab:CSF}).

The unbiased  mass ratio distribution for visual  binaries with B-type
primaries is the  main result of this study.   The idea of independent
selection of the visual components from some initial mass function can
now  be  definitely rejected.   However,  the  new  result is  not  so
unexpected,  after all.   The $f(q)$  obtained  by DM91  for the  wide
($\log P  ({\rm days}) >  4$) systems is  similar to the  $f(q)$ found
here (Fig.~\ref{fig:qhist}).   The mass ratio should  be indeed biased
towards a uniform  one by stellar dynamics, whatever  the IMF.  N-body
simulations  demonstrated that  the  shape of  $f(q)$  depends on  the
density and  composition of the  stellar aggregate where  the binaries
have been formed, and a certain choice of parameters may reproduce the
result    of     DM91    (Kroupa    \cite{Kroupa95},     Durisen    et
al. \cite{Durisen01}).

On the other hand, the  secondary components of B-type stars must have
been formed in the same  clouds as their primaries, in conditions that
likely favored high masses. The  secondaries may thus be distinct from
the  rest of  low-mass population  in Sco  OB2 with  respect  to their
initial mass  function and  age.  For  the moment we  are not  able to
disentangle the influence of dynamical  and birth factors on the final
mass ratio distribution. The  important thing is that the distribution
itself is now known with some confidence.

\begin{acknowledgements}

The  Fellowship of  the Belgian  {\em Services  Federaux  des Affaires
Scientifiques,  Techniques and  Culturelles} provided  the possibility
for N.S.   to work at the  Royal Observatory of  Belgium.  Authors are
grateful to the  staff of ESO 3.6m telescope,  especially to O.~Marco,
for  their support of  observations.  Thanks  to A.~Chalabaev  for his
help with data acquisition and stimulating discussions.  This research
made use of Simbad database operated at CDS, Strasbourg, France and of
the  Digital  Sky  Survey  produced  at the  Space  Telescope  Science
Institute, USA.

\end{acknowledgements}

\appendix

\section{Detection and measurement of close companions with DAOPHOT}
\label{app:dao}

\begin{figure}
\psfig{figure=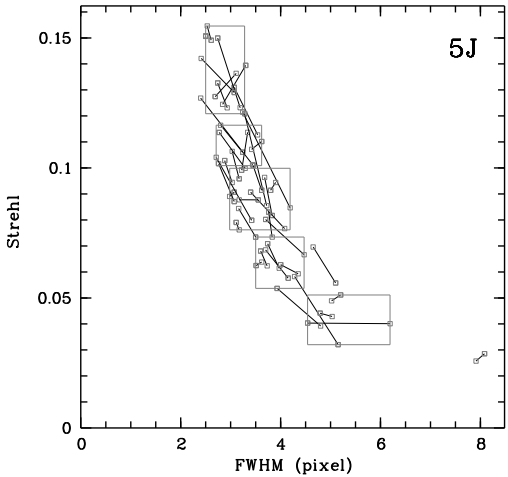,width=7cm,clip=t}
\caption{\label{fig:psf} Example of  the classification of images into
groups by Strehl  ratio (SR), used to produce average  PSFs. The SR of
images of the last (fifth) night in $J$ band are plotted against their
Full Width at Half Maximum  (FWHM); boxes denote the borders of groups.
Lines connect the  points which belong to two planes  of the same data
cube and reflect the fast variability of image quality.  }
\end{figure}

In  this  appendix we  present  the solution  of  the  problem of  PSF
selection  for DAOPHOT  fitting in  non-coronographic mode.   A simple
subtraction of the radial profile of PSF ({\tt jupe} algorithm) leaves
the  bright  semi-static  speckle   pattern  around  the  target  star
unattenuated.  To  remove it  partially, we produced  the set  of {\em
average} PSFs for each night and  each filter, grouped by the value of
the   Strehl   ratio  (i.e.    relative   sharpness)   of  the   image
(Fig.~\ref{fig:psf}).  Being  averaged over  images  of many  objects,
these synthetic PSFs contain no trace of any possible faint companions
which  hide  under  the  speckle  pattern ({\em  extra  PSF  cleaning}
function of  DAOPHOT II removes  all outlying features like
median filtering).   The radius of  these synthetic PSFs is  35 pixels
$=1\farcs75$ (maximal  available in DAOPHOT  from the used  release of
ESO-MIDAS).

All non-coronographic images  of target stars were fitted  with one of
those average PSFs selected according  to their own Strehl ratio.  The
residuals after  PSF suntraction were visually searched  for new close
companions.    The   companions   of   all   new   and   known   close
($\rho<1\farcs75$) visual systems were simultaneously fitted with this
method  using {\tt  NSTAR}  or  {\tt ALLSTAR}  utility  of DAOPHOT  to
produce  the  differential  astrometric and  photometric  measurements
reported in this paper.

To get an  estimate of the detection limits  achieved with subtraction
of  synthetic  PSFs, we  used  again  the  {\tt jupe}  program.  After
subtraction, the intensity of the remaining speckle noise decreased to
a  level which,  by  chance,  coincided with  the  detection limit in
coronographic images (Fig.~\ref{fig:detlim}).

\section{Differential astrometry with coronograph}
\label{app:ast}

The coordinate  differences between the  source and target  stars were
measured in three different ways.   The first and most reliable method
is  the simultaneous  PSF fitting  to primary  and secondary  stars in
non-coronographic images.

However, simultaneous  PSF fitting is not  applicable to coronographic
images since the  primary star is not visible.   Instead, the position
of the  primary was determined  from the PSF  wings by the  {\tt jupe}
program.  The relation between this  method and direct PSF fitting was
studied.  The precision  of {\tt  jupe} coordinates  of a  primary was
found  to be  $\pm 0.3$  pixels, with  a constant  bias of  $-1.5$ and
$-0.6$ pixels in  $x$ and $y$ directions, respectively.   This bias is
caused, possibly, by the asymmetry of the PSF wings.

Alternatively, the known positions of the primary in the two quadrants
of non-coronographic images can be used to predict the position of the
star under the mask, supposing  that AO system stabilizes the image in
the  detector plane  and that  the  offsets provided  by the  chopping
mirror  of ADONIS  are  precise and  repeatable.   These offsets  were
studied  and calibrated.  It  turned out  that the  chopping mechanism
moves the target star across the detector with an rms error of about $
\pm 0.25$ pixels;  the error never exceeds 0.6 pixels.

Whenever  components  were  measured  by  several  methods,  resulting
coordinates  were computed as  weighted averages,  with weight  10 for
direct PSF fitting,  weight 1.5 for {\tt jupe}  coordinates and weight
2.0 for coordinates extrapolated from non-coronographic images.

\end{document}